\documentclass[aps,prx,twocolumn,longbibliography,floatfix,10pt,superscriptaddress]{revtex4-1}
\usepackage[utf8]{inputenc}

\usepackage{natbib}
\usepackage{graphicx}
\usepackage{latexsym}
\usepackage{amssymb}
\usepackage{amsmath}
\usepackage{amsfonts}
\usepackage{gensymb}
\usepackage{bm}
\usepackage{hyperref}
\usepackage{braket}
\usepackage{physics}
\usepackage{bbold}
\usepackage[caption=false]{subfig}
\usepackage[dvipsnames]{xcolor}
\usepackage[normalem]{ulem}
\usepackage{siunitx}
\usepackage{mhchem}
\usepackage{multirow}

\newcommand{\PRLsec}[1]{\emph{#1---}}

\newcommand{\supplementarysection}{%
  \setcounter{figure}{0}
  \let\oldthefigure\thefigure
  \renewcommand{\thefigure}{S\oldthefigure}
  \setcounter{section}{0}
  \let\oldthesection\thesection
  \renewcommand{\thesection}{S\oldthesection}
  \setcounter{equation}{0}
  \let\oldtheequation\theequation
  \renewcommand{\theequation}{S\oldtheequation}
  \setcounter{table}{0}
  \let\oldthetable\thetable
  \renewcommand{\thetable}{S\oldthetable}
}

\begin{document}

\title{Topology, magnetism and charge order in twisted \ce{MoTe2} at higher integer hole fillings}

\author{Taige Wang}
\thanks{These authors contributed equally.}
\affiliation{Department of Physics, University of California, Berkeley, CA 94720, USA \looseness=-1}
\affiliation{Material Science Division, Lawrence Berkeley National Laboratory, Berkeley, CA 94720, USA \looseness=-1}

\author{Minxuan Wang}
\thanks{These authors contributed equally.}
\affiliation{Department of Physics, University of California, Berkeley, CA 94720, USA \looseness=-1}

\author{Woochang Kim}
\thanks{These authors contributed equally.}
\affiliation{Department of Physics, University of California, Berkeley, CA 94720, USA \looseness=-1}
\affiliation{Material Science Division, Lawrence Berkeley National Laboratory, Berkeley, CA 94720, USA \looseness=-1}

\author{Steven G. Louie}
\affiliation{Department of Physics, University of California, Berkeley, CA 94720, USA \looseness=-1}
\affiliation{Material Science Division, Lawrence Berkeley National Laboratory, Berkeley, CA 94720, USA \looseness=-1}

\author{Liang Fu}
\affiliation{Department of Physics, Massachusetts Institute of Technology, Cambridge, MA 02139 USA \looseness=-1}

\author{Michael P. Zaletel}
\affiliation{Department of Physics, University of California, Berkeley, CA 94720, USA \looseness=-1}
\affiliation{Material Science Division, Lawrence Berkeley National Laboratory, Berkeley, CA 94720, USA \looseness=-1}

\date{\today}

\begin{abstract}

Twisted homobilayer transition metal dichalcogenide (TMD) attracts an expanding experimental interest recently for exhibiting a variety of topological and magnetic states even at zero magnetic field. Most of the studies right now focus on hole filling $\nu_h \leq 1$, while a rich phase diagram at higher hole filling calls for more investigation. We perform a thorough survey of possible interaction-driven phases at higher integer hole fillings. We first construct the continuum model from a first-principles calculation, and then perform a self-consistent Hartree-Fock study of the interacting ground states. We identify various valley polarized (VP) states at odd integer fillings and intervalley coherent (IVC) states at even integer fillings and discuss the energetics competition among them. We also discuss the origin and the experimental implications of the curious Chern insulator at $\nu_h = 2$.

\end{abstract}

\maketitle

Two-dimensional moir\'e superlattices have proven to host rich physics including superconductivity, ferromagnetism, and topological order. Very recently, twisted homobilayer transition metal dichalcogenides (TMD) have attracted great deal of attention both theoretically and experimentally. 
In particular, twisted homobilayer \ce{MoTe2} exhibits robust ferromagnetism and quantum anomalous Hall effects, as evidenced by the high transition temperatures up to \SI{14}{\K} \cite{anderson_programming_2023, cai_signatures_2023, zeng_integer_2023, park_observation_2023,  xu2023observation}. Even more surprisingly, the first transport evidence for  fractional quantum anomalous Hall states (FQAH), i.e., fractional Chern insulators arising from the spontaneous breaking of time-reversal symmetry, were found at fractional moir\'e fillings of twisted \ce{MoTe2} \cite{park2023observation,xu2023observation}. The observed ferromagnetism and quantized anomalous Hall effects, as theoretically anticipated \cite{LiangNC,li_spontaneous_2021,crepel_anomalous_2023}, both arise   
from Coulomb interaction in a topological moir\'e band of Kane-Mele type \cite{wu_topological_2019, morales-duran_pressure-enhanced_2023, wang_fractional_2023, reddy_fractional_2023, yu_fractional_2023,wang_topological_2023,goldman_zero-field_2023,reddy_toward_2023,dong_composite_2023,morales-duran_magic_2023}.    

So far most of the experimental efforts have been focused on hole filling $\nu_h \leq 1$, where correlated topological states are found over a remarkably wide range of twisted angles up to at least $\sim 4 \degree$ \cite{anderson_programming_2023,zeng_integer_2023}. At these twist angles, the bandwidth of the first few occupied moir\'e bands are actually comparable \cite{LiangNC}. One may thus expect strong interaction effect and correlated states to exist at higher hole fillings as well. A few recent theoretical works using effective models have predicted various topological states at hole-filling $\nu_h = 2$ \cite{Xiaonu2,Fengchengnu2,FengchengMajorana}. However, an accurate and systematic study of interaction induced phases at integer fillings $\nu_h \geq 2$ is still lacking due to the incomplete understanding of the electronic structure in twisted \ce{MoTe2}. 
As a good starting point, a reliable study at higher fillings requires an accurate continuum model that correctly captures moir\'e bands beyond the first one. There have been a few first principle studies on twisted \ce{MoTe2} with various methods and approximations. Differing results on the band structure and topology, especially concerning the second valence band, have been reported \cite{wang_fractional_2023,reddy_fractional_2023}. 

In this work, we perform a large-scale DFT calculation of twisted \ce{MoTe2} at $\theta = 3.89 \degree$ with full lattice relaxation. The continuum model we derive from the study reproduce not only the DFT band structure but also the charge density from the DFT calculation. 
To map out the ground state phase diagram of twisted \ce{MoTe2} at integer fillings, we perform a self-consistent Hartree-Fock (HF) calculation based on the continuum model obtained from DFT with Coulomb interactions. 
We find that generally, the valley polarized (VP) phases---which exhibit ferromagnetic order---dominate at odd fillings, while the intervalley coherent phases (IVC)---which exhibit antiferromagnetic order---dominate at even fillings. 
We provide an intuitive understanding of the competition between VP and IVC states based on the Chern number structure of the first few bands. At $\nu_h = 2$, our phase diagram mostly agrees with previous studies \cite{Xiaonu2,Fengchengnu2,FengchengMajorana}. We present a simple real-space understanding of the curious antiferromagnetic Chern insulator and reveal the charge order associated with it \cite{Xiaonu2,FengchengMajorana,ZiqiangAFMCI}. At $\nu_h = 3$, we identify a few partially valley polarized (PVP) states.  

\begin{figure}[htbp]
    \centering
    \includegraphics[width=0.48\textwidth]{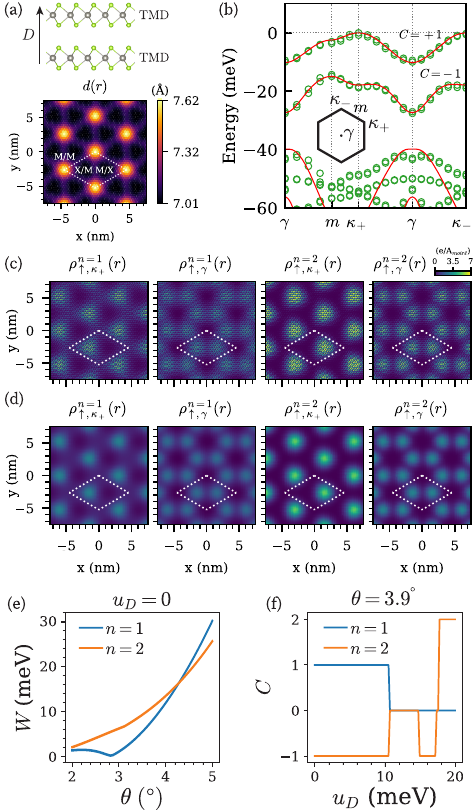}
    \caption{(a) Atomic structure and the relaxed interlayer distance distribution in the $3.89 \degree$ twisted moir\'e superlattice.  (b) Single-particle band structure at twist angle $\theta = 3.89^{\degree}$ and zero displacement field from first-principles calculation (green circles) and the continuum model fitting (red lines). We also label the valley Chern number of the first two bands in the $K$ valley (spin up sector). (c-d) charge density of the Bloch wavefunctions at high symmetry points in the first two valance bands from (c) DFT calculation and (d) continuum model calculation. (e-f) Bandwidths and Chern numbers of the first two bands (from $K$ valley) as a function of the twist angle $\theta$ and the applied displacement field $u_D$.}
    \label{fig:DFT}    
\end{figure}

\PRLsec{First principles DFT calculation} We construct a moir\'e superlattice using two monolayers of \ce{MoTe2} at a twisted angle of $3.89 \degree$. The single-layer unit-cell lattice constant was determined via structural relaxation from DFT calculations employing the Perdew-Burke-Ernzerhof-type (PBE) generalized gradient approximation \cite{GGA}. The monolayer's unit-cell lattice constant, as calculated in this study, is \SI{3.55}{\angstrom}. Subsequently, upon constructing the rigidly twisted bilayer, we relaxed the  moir\'e superlattice structure by minimizing the system's total energy using the PBE-D2 method \cite{GGAD2}, which accounts for van der Waals interactions. All relaxation calculations were conducted using a plane wave basis set and optimized norm-conserving Vanderbilt pseudopotentials \cite{ONCVPP, PseudoDojo}, implemented within the {\tt Quantum ESPRESSO} package \cite{QE} (see Ref.~\onlinecite{SM} for details).

Figure~\ref{fig:DFT}(a) shows the distribution of interlayer distances $d(r)$ in the fully relaxed moir\'e superlattice (defined as the separation between Mo atoms of the two layers locally). The relaxation pattern aligns well with previous studies conducted on homobilayer TMD structures \cite{2018_Mit, 2022_Kundu, wang_fractional_2023,reddy_fractional_2023}. The interlayer separation ranges between $\SI{7.01}{}-\SI{7.62}{\angstrom}$, which are larger than those reported in Ref.~\onlinecite{wang_fractional_2023}.

From the relaxed atomic structure, we performed DFT calculations to obtain its Kohn-Sham orbital band structure, employing the {\tt SIESTA} \cite{SIESTA} code. These calculations used double-zeta-polarization (DZP) basis sets, successfully capturing the spectrum of bilayer \ce{MoTe2} systems (refer to Ref.~\cite{SM}). Figure~\ref{fig:DFT}(b) presents the DFT band structures of twisted bilayer \ce{MoTe2} at a twisted angle of $3.89 \degree$. Analysis of our calculations reveals that the first three moir\'e valance bands originate from the $K$ and $K'$ valleys of the underlying MoTe$_2$ layers. The bandwidths of the first and second valence bands are \SI{10.0}{} and \SI{13.1}{} 
meV, respectively. Additionally, the moir\'e bands from the $\Gamma$-valley are situated approximately \SI{50}{meV} below the valence band maxima (VBM). It's noteworthy that the relative energy difference between the $\Gamma$ and $K$ valley states is highly sensitive to the atomic-scale lattice structure. Previous studies have revealed that when a smaller isolated single-layer lattice constant is used, the $\Gamma$-valley bands are further separated from the low-energy $K/K'$-valley bands of interest \cite{2022_Kundu}. 

\begin{table}[h!]
\renewcommand*{\arraystretch}{1.3}
\begin{tabular*}{.35\textwidth}{@{\extracolsep{\fill}}ccccc}
\hline \hline
\textrm{band} &
\textrm{$\kappa_{+}$} &
\textrm{$\gamma$} &
\textrm{$\kappa_{-}$} &
\textrm{$C$} \\
\hline
$1$ & $\omega$ & $-1$ & $\omega$ & $+1$ \\
$2$ & $\omega^*$ & $-1$ & $\omega^*$ & $-1$ \\
\hline \hline
\end{tabular*}
\caption{$C_3$ eigenvalues and the corresponding valley Chern numbers for the first two valence bands (from $K$-valley, spin-up) electrons, where $\omega = e^{i\pi/3}$.}
\label{tab:dftc3}
\end{table}

To study the topological aspects of the band structure, we computed the $C_3$ eigenvalues of the first and second valence bands for a given spin channel at the $C_3$ symmetric momenta, $\gamma, \kappa_-, \kappa_+$. 
These symmetry eigenvalues determine the spin Chern number mod $3$, as shown for twisted TMD homobilayers in Ref.\cite{LiangNC}. 
Table~\ref{tab:dftc3} presents the $C_3$ eigenvalues of the first and second valence bands from our DFT calculation, which are well-separated from each other and from other bands. The computed $C_3$ values indicate that the first and second valence bands derived from $K$ valley (spin up) electrons have opposite valley Chern numbers: $+1$ and $-1$, respectively.

\PRLsec{Continuum model} Following the formulation developed in Ref.~\cite{wu_topological_2019, reddy_fractional_2023}, we write down a continuum model Hamiltonian,
\begin{equation} \label{eq:hamiltonian}
\begin{gathered}
    H_0=\sum_{\tau, l, \mathbf{r}} c_{\tau, l, \mathbf{r}}^{\dagger}\left([h_{\tau}]_{ll'}-\mu \delta_{l l'}\right) c_{\tau, l', \mathbf{r}}\\
    [h_{K}]_{ll'}=\left(\begin{array}{cc}h^K_+ + V_{+}(\boldsymbol{r}) + u_D & T(\boldsymbol{r}) \\ T^{\dagger}(\boldsymbol{r}) & h^K_- + V_{-}(\boldsymbol{r}) - u_D\end{array}\right)
\end{gathered}
\end{equation}
where $\tau$ is the valley index and $l = \pm$ is the layer index. Here the spin degree of freedom is locked to the valley degree of freedom ($K$/$K'$) due to Ising spin-orbit coupling. The kinetic part $h^K_{\pm}$ in the $K$ valley takes the form
\begin{equation} \label{eq:potential}
    h^K_{\pm} = -\frac{\hbar^2\left(-i \nabla-\kappa_{\pm}\right)^2}{2 m^*}
\end{equation}
where $m^*$ is the effective mass. The moir\'e structure results in an intralayer moir\'e potential $V(r)$ and an interlayer tunneling $T(r)$,
\begin{equation}
\begin{gathered}
    V_{\pm}(\mathbf{r})=2 V \sum_{j=1,3,5} \cos \left(\mathbf{g}_j \cdot \mathbf{r} \pm \phi \right),\\
    T(\mathbf{r})=w \left(1+e^{-i \mathbf{g}_2 \cdot \mathbf{r}}+e^{-i \mathbf{g}_3 \cdot \mathbf{r}}\right)
\end{gathered}
\end{equation}
where $\mathbf{g}_j$ are the moir\'e reciprocal lattice vectors.

Here $m^*$, $V$, $\phi$ and $w$ are all model parameters that need to be extracted from the first-principles calculation. We limit ourselves to the first harmonics (i.e., the first shell of $\mathbf{g}$ vectors) so that the fitting procedure is more controlled. Fitting the continuum model band structure to both the first and second DFT bands as shown in Fig.~\ref{fig:DFT}(b), we obtain the parameters in Table~\ref{tab:parameter}. To verify the validity of the continuum model, we first compute the Chern numbers of the first two bands at $\theta = 3.89 \degree$ (see Fig.~\ref{fig:DFT}(f)), which agree with the Chern numbers extracted by the $C_3$ eigenvalues discussed above. 

To further demonstrate the accuracy of our continuum model,  we present a direct comparison of the charge density obtained from the continuum model and the DFT calculation (see Ref.~\onlinecite{SM} for details). 
In Fig.~\ref{fig:DFT}(c-d), we show the electron density $\rho^{n}_{\uparrow, k}(r) =  |\psi_{\uparrow, k}^{n}(r)|^2$ of the Bloch wavefunction at momentum $k$ and band $n$. The continuum model not only reproduces the triangular lattice pattern, but also agrees quantitatively well with the DFT calculation \cite{SM}.

\begin{table}[htbp]
    \centering
    \renewcommand*{\arraystretch}{1.3}
    \begin{tabular*}{.48\textwidth}{@{\extracolsep{\fill}}cccccc}
    \hline \hline  & $\phi \;(\degree)$ & $V \;$(meV) & $w \;$(meV) & $m^* \;(m_e)$ & $a_0 \;(\AA)$ \\
    \hline Ref.~\onlinecite{reddy2023fractional} & $-100$ & $  7.5$ & $-11.3$ & $0.62$ & $3.52$ \\
     Ref.~\onlinecite{wang_fractional_2023} & $-107.7$ & $  20.8$ & $-23.8$ & $0.6$ & $3.52$ \\
    This work & $-107.7$ & $17.0$ & $-16.0$ & $0.62$ & $3.55$ \\
    \hline \hline
    \end{tabular*}
    \caption{Continuum model parameters of twisted \ce{MoTe2}. In all HF calculations we adopt our own parameters in the last row.}
    \label{tab:parameter}
\end{table}

Using the continuum model parameters extracted from our DFT band structure at $\theta=3.89 \degree$, we calculate the  band structures of twisted \ce{MoTe2} at various twist angles. Both the bandwidth and topology of moir\'e bands play an important role in the phase diagram. 
While early theoretical studies have largely focused on the magic angle where the band is almost completely flat, 
the quantum anomalous Hall state at $\nu_h=1$ has been experimentally observed up to at least $4 \degree$. 
At around $4 \degree$, we note that both first and second bands have a similar bandwidth of around $\SI{10}{\meV}$, as shown in Fig.~\ref{fig:DFT}(e). This motivates us to explore interaction induced phases at higher integer fillings $\nu_h = 2, 3$. We also point out that 
our continuum model parameters result in a magic angle $\theta_m \sim 2.8 \degree$ at which the first band becomes extremely flat. 

We further discuss band topology.   
The skyrmion texture of the layer pseudospin gives the first valance band valley Chern number $C_K = +1$ in twisted \ce{MoTe2} \cite{reddy_fractional_2023}. The valley Chern number of the second band, however, depends sensitively on the continuum model parameters. Previously proposed continuum model parameters result in a sign change of the Chern number of the second band 
as the twist angle increases \cite{reddy_fractional_2023}. This transition, however, is not seen within our continuum model, suggesting a large gap between the second and the third band. At sufficiently large displacement field $u_D \sim \SI{10}{\meV}$, the electrons are polarized to the triangular lattice of one of the layers and then the Chern number vanishes (see Fig.~\ref{fig:DFT}(f)).

\begin{figure}[tbp]
    \centering
    \includegraphics[width = 0.48\textwidth]{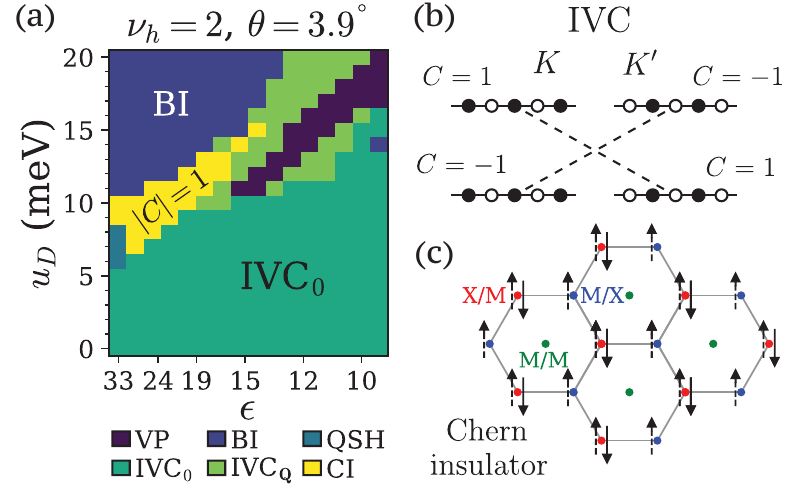}
    \caption{(a) Interacting phase diagram at $\nu_h = 2$ of twisted \ce{MoTe2} as a function of displacement field and dielectric constant at twist angle $\theta = 3.9^{\degree}$. Different colors represent different phases (see main text for definition). (b) Schematic diagram of the IVC state at $\nu_h = 2$, which hybridizes between bands at different energies but with the same Chern number. (c) Schematic diagram of the Chern insulator. Red (XM) and blue (MX) dots represent the triangular lattice of the bottom and top layer. Dashed arrows indicate $1/2$ occupation. Here the $K$ valley (spin up) holes are sublattice unpolarized, whereas the $K'$ valley (spin down) holes are sublattice polarized, exhibiting a non-trivial charge order between the XM and MX region.}
    \label{fig:nu2}    
\end{figure}

\PRLsec{$\nu_h = 2$ phase diagram} To understand possible correlated states at higher integer fillings, we perform self-consistent Hartree-Fock (SCHF) calculations on a $30 \times 30$ momentum grid with screened Coulomb interaction projected to the top five valence bands per valley (see Ref.~\onlinecite{SM} for details). We start by looking at filling two holes per moir\'e unit cell $\nu_h = 2$. We present the rich interacting phase diagram in Fig.~\ref{fig:nu2} (a), which agrees qualitatively with previous HF studies at $\nu_h = 2$ \cite{Xiaonu2,Fengchengnu2,FengchengMajorana}. The competing phases include a trivial band insulator (BI), a quantum spin Hall (QSH) insulator, a valley polarized (VP) insulator, two different types of intervalley coherent (IVC) states at mini-BZ momentum $0$ and momentum $\mathbf{Q} = \kappa_+ - \kappa_- = \kappa_-$, and a valley unpolarized Chern insulator (CI) with Chern number $|C| = 1$. 

Due to spin-valley locking, valley order can also be interpreted as spin order. For example, the VP state is a simple ferromagnet, filling the first two hole bands from the $K$ valley (or equivalently the spin up sector); in Tab.\ref{tab:correspondence} we translate to the spin nomenclature used in previous works \cite{Xiaonu2,Fengchengnu2,FengchengMajorana,ZiqiangAFMCI}. 
Both the QSH and BI states arises from filling the first hole band from both valleys (spins).
However, within the single-particle band structure the bands undergo a Chern-changing transition at a critical $u_D$ (see Fig.~\ref{fig:DFT} (f)) : for low $u_D$, they have opposing $C = \pm 1$, leading to QSH, while for large $u_D$ they have $C=0$, giving the trivial BI. 
The QSH region is relatively small within our continuum model due to the relatively low critical displacement field $u_D \sim \SI{10}{meV}$ (see Fig.~\ref{fig:DFT} (f)). Both IVC states can be interpreted as simultaneous intralayer antiferromagnets and interlayer antiferromagnets (see Ref.~\cite{SM} for the spin texture). In particular, the IVC with momentum $\mathbf{Q}$ is a $120 \degree$ antiferromagnet, or equivalently a $\sqrt{3} \times \sqrt{3}$ spin density wave, at the moir\'e scale \cite{wang_magnon}. While these IVC states break the physical time-reversal symmetry $\mathcal{T} = i \tau_y K$ satisfying $\mathcal{T}^2 = -1$, they still respect the non-Kramers time-reversal symmetry $\mathcal{T}^{\prime}=e^{i \pi \tau^z / 2} \mathcal{T} = \tau_x K$ satisfying ${\mathcal{T}^{\prime}}^2 = 1$ and therefore their Chern numbers vanish (here $\tau$'s are the Pauli matrices in the valley space, and $K$ is the complex conjugation).

\begin{table}[htbp]
    \centering
    \renewcommand*{\arraystretch}{1.3}
    \begin{tabular*}{.4\textwidth}{@{\extracolsep{\fill}}ccc}
    \hline \hline \multicolumn{2}{c}{Magnetic orders} & \multirow{2}{*}{Valley-ordered phases} \\
    Intralayer & Interlayer & \\
    \hline 
    FM$_z$ & AFM$_z$ & CI\\
    FM$_z$ & FM$_z$ & VP\\
    AFM$_z$ & AFM$_z$ & IVC$_{\mathbf{0}}$\\
    $120 \degree$ AFM & AFM$_{xy}$ & IVC$_{\mathbf{Q}}$\\
    \hline \hline
    \end{tabular*}
    \caption{Topology and magnetic order of various interacting ground states at $\nu_h = 2$. Previous literature labels these states with their interlayer magnetic orders \cite{Xiaonu2,Fengchengnu2,FengchengMajorana,ZiqiangAFMCI}.}
    \label{tab:correspondence}
\end{table}

Compared to $\nu_h = 1, 3$ (see next section), at $\nu_h = 2$ the IVC states dominate the phase diagram at low $u_D$. This behavior can be readily understood from the Chern number structure of the first two hole bands. Within our continuum model, the lowest two bands carry opposite Chern number (see Fig.~\ref{fig:DFT} (f)). At $\nu_h = 2$, the IVC state always hybridizes between bands at different energies but with the same Chern number as shown in Fig.~\ref{fig:nu2} (b). Same-Chern hybridization avoids the exchange-energy penalty which arises when hybridizing bands with opposite Chern number \cite{BCZ2020,DGGmoire,Serlin2019,ZMS2019,NickPRX,TaigeNC}. At odd integer fillings, hybridizing bands at different energies are strongly penalized by the kinetic energy cost, and therefore VP states dominate instead.

\begin{figure*}[htbp]
    \centering
    \includegraphics[width = 0.78\textwidth]{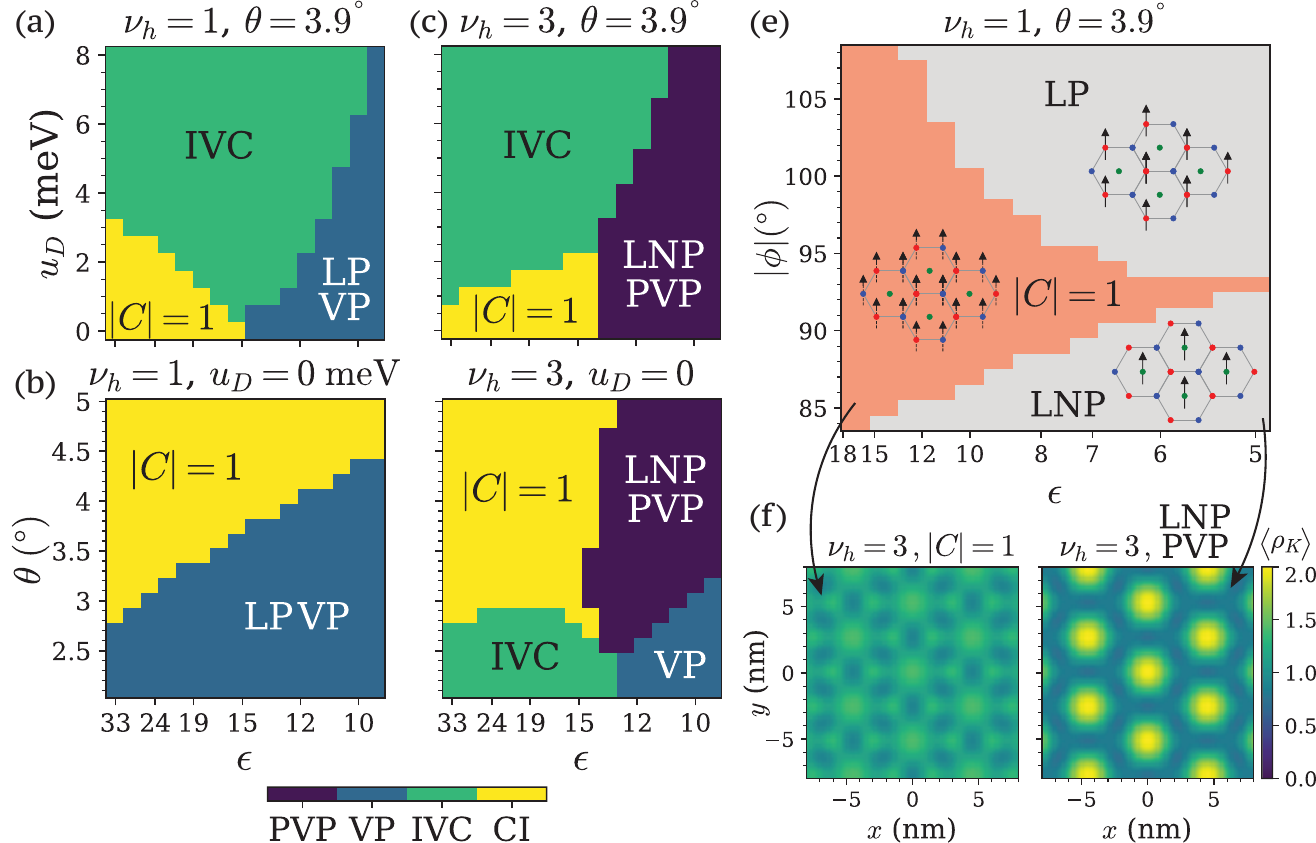}
    \caption{(a) Interacting phase diagram at $\nu_h = 1$ of twisted \ce{MoTe2} as a function of displacement field $u_D$ and dielectric constant $\epsilon$ at twist angle $\theta = 3.9^{\degree}$. Different colors represent different phases (see main text for definition). (b) Interacting phase diagram as a function of twist angle and dielectric constant at zero displacement field. (c-d) Similar phase diagrams at $\nu_h = 3$, resembling the $\nu_h = 1$ phase diagram. (e) Typical interacting phase diagram at $\nu_h = 1$ assuming full valley polarization. Here we fix $u_D = 0$ and sweep the phase $\phi$ of the moir\'e potential. Schematics of each state are shown as insets. (f) Typical charge density distribution of the $K$ valley electrons in the LNP-PVP ($u_D = 0$, $\epsilon = 12.5$) phase and the CI ($u_D = 0$, $\epsilon = 16.7$) phase at $\nu_h = 3$, which can be identified with the LNP and CI phase at $\nu_h = 1$.}
    \label{fig:nu3}    
\end{figure*}

We also find a curious $|C|=1$ Chern insulator which breaks the time-reversal symmetries but has vanishing valley (spin) polarization. This state was first proposed in the Kane-Mele-Hubbard model as an AFM-CI state \cite{ZiqiangAFMCI}, and later studied in an HF calculation of twisted \ce{MoTe2} \cite{Xiaonu2, FengchengMajorana}. The Chern insulator of interest always appear above the critical $u_D$ where the electrons are expected to polarize to the bottom layer and form a trivial BI on a triangular lattice of the XM regions. Now if we turn on Coulomb repulsion, some of the spin up electrons will be transferred to the MX regions to reduce correlation energy, forming a Haldane insulator on a honeycomb lattice from both layers similar to the $u_D = 0$ limit; while the spin down electrons remain polarized to the XM regions to minimize exchange energy, forming a trivial BI on a triangular lattice from the bottom layer (see Fig.~\ref{fig:nu2} (c)). The appearance of this Chern insulator relies on a delicate balance between the displacement field and the Coulomb repulsion: when the displacement field dominates, both spin species will be polarized to the bottom layer and form a trivial BI; when Coulomb repulsion dominates, any sublattice polarization is unfavorable, then the QSH insulator and certain IVC order will appear. This simple picture not only explains the origin of this Chern insulator but also predicts a particular charge order associated with it: the charge density at the XM and the MX region should be almost quantized to $\rho^{\mathrm{XM}}/\rho^{\mathrm{MX}} = 3$ when the interaction is strong, which allows future charge sensing STM experiments to uniquely identify this state \cite{FengSTM}. 

\PRLsec{$\nu_h = 3$ phase diagram} As shown in Fig.~\ref{fig:nu3} (a-d), the $\nu_h = 3$ phase diagram qualitatively resembles the $\nu_h = 1$ phase diagram. The competing states include an IVC state and two types of partially valley polarized (PVP) states, one with Chern number $|C| = 1$ and one without. We will postpone our discussion of the IVC phase and focus on the competition between the two PVP states for now. Both the $\nu_h = 1$ and the $\nu_h = 3$ phase diagram feature a topological transition at some critical interaction strength. Such resemblance can be understood from a simple picture: in the PVP phase, the first two bands in the $K'$ valley are trivially filled and serve as a dielectric background for the $K$ valley holes at $\nu_h^K = 1$. However, on closer examination, the trivial VP state at $\nu_h = 1$ has finite layer polarization while the trivial PVP state at $\nu_h = 3$ does not.

Before diving into the complicated PVP phases, we first examine possible VP phases at $\nu_h = 1$. As shown in Fig.~\ref{fig:nu3} (e), three possible VP phases at $\nu_h = 1$ are a layer polarized (LP) trivial state, a topological state (CI) with $|C| = 1$, and a layer-unpolarized (LNP) trivial state \cite{wang_magnon, reddy_fractional_2023, Multiferroicity}. At low interaction strength, one always gets a topological state, and the interaction will drive a transition into either an LP state or an LNP state depending on the moir\'e potential. The competition of the LP and LNP states can be understood from the real space configuration in Fig.~\ref{fig:nu3} (e) insets: the phase of the moir\'e potential determines whether the potential minimum occurs at the XM/MX region or the MM region, which in turn favors the LP state or the LNP state respectively.

At $\nu_h = 3$, the two filled bands in the $K'$ valley strongly renormalize the effective moir\'e potential in the $K$ valley. At HF level, the additional piece in the effective potential comes from the Hartree potential,
\begin{equation}
    V_{\mathrm{eff}}^K(\mathbf{r}) = V^K(\mathbf{r}) + \int \dd \mathbf{r}'  V_C(\mathbf{r}-\mathbf{r}') \rho_{K'}(\mathbf{r}')
\end{equation}
where $\rho_{K'}(\mathbf{r})$ is the charge density of the $K'$ valley (spin down) holes, and $V_C$ is the screened Coulomb repulsion. Now we check the $K$ valley charge density of the topological and trivial PVP states at $\nu_h = 3$ and indeed find an excellent agreement with the real space picture we developed for the topological and LNP state at $\nu_h = 1$ (see Fig.~\ref{fig:nu3} (f)). In other words, going to filling $\nu_h = 3$ allows us to access parameter regimes that is inaccessible at $\nu_h = 1$.

Now we return to the IVC states at $\nu_h = 3$. At $\theta > 3 \degree$ and large $u_D$, the IVC state can be simply viewed as a coherent superposition of the second band from two valleys, with the first bands filled as a dielectric background. Therefore, it can still be viewed as a second band analog of the IVC state at $\nu_h = 1$. Similarly, it benefits from the more dispersive bands at large $u_D$ by saving the kinetic energy when canting the valley isospin vector in momentum space  \cite{AdrianTBG,EslamNC,TaigeNC}. The IVC state at $\theta < 3 \degree$ is more complicated. The bandwidth there is so small such that mixing the first four bands from both valleys in total becomes possible (see Fig.~\ref{fig:DFT} (e)). The IVC state chooses to occupy a particular superposition of the four bands in a time reversal $\mathcal{T}'$ symmetric way.

\PRLsec{Discussion} Our survey of possible ground states at higher integer fillings establish certain expectations for future experiments on twisted \ce{MoTe2}. The rich phase diagram we obtained suggest that twisted \ce{MoTe2} at higher integer fillings is a versatile platform for topological states, magnetism. In particular, it further supports charge order that is believed to be absent at hole filling $\nu_h = 1$ \cite{LiangNC}. We also discuss the experimental implication of these states, whose topological, magnetic and charge order nature makes it possible to detect them via transport, magnetic circular dichroism (MCD), and charge sensing STM techniques respectively.

As a indispensable step of our study, we construct a continuum model from a large-scale brute-force DFT study on twisted \ce{MoTe2} with full lattice relaxation. We carefully match the charge distribution to ensure our continuum model not only cpatures the band structure but also structures of the Bloch wavefunction. From the $C_3$ symmetry eigenvalues we pin down the opposite Chern number of the first two bands, which supports the Kane-Mele modeling of twisted \ce{MoTe2} \cite{LiangNC,LiangPRB,Xiaonu2}. We also identify an magic angle close to $3 \degree$, which can be one of the hints to various interacting physics around $3 \degree$ observed in the experiment \cite{anderson_programming_2023, cai_signatures_2023, zeng_integer_2023, park_observation_2023,  xu2023observation}.

Instead of focusing on a particular filling, we discuss the general structure at even vs odd integer fillings. The VP states always dominate at odd fillings, while the IVC states take over at even fillings. This behavior can be understood from the Chern number structure of the single particle bands. At even fillings, IVC states are allowed to hybridize between bands with different energy but same Chern number, so that they can avoid the exchange energy penalty when hybridizing between bands with opposite Chern number. This alternating pattern makes $\nu_h = 2$ an excellent platform for studying IVC physics given the relatively large displacement field required to access the IVC phase at $\nu_h = 1$. 

Though the curious Chern insulator at $\nu_h = 2$ has been studied previously, the focus has been on the topological point of view, especially the Weyl points between the band insulator (BI) and the QSH insulator \cite{Xiaonu2,FengchengMajorana,ZiqiangAFMCI}. We present a simple real space picture to understand its origin, as one of the spin species is polarized to one sublattice at finite displacement field, the competition between the displacement field and the Coulomb repulsion will force the other spin specie to stay on the honeycomb lattice from both layers, resulting a Haldane insulator. The nearby BI and QSH states are simply two limits where the displacement field dominates over the Coulomb repulsion or vice versa. This real space picture also associates this Chern insulator with a particular charge order that can be detcted by future STM experiments.

{\em Note added.} Towards the completion of the project, we become aware of three related DFT calculations of twisted homobilayer \ce{MoTe2}, which qualitatively agree with our results \cite{BernevigDFT, YangDFT, XiaoDFT}.

\begin{acknowledgments}
We thank Xiaodong Xu, Shubhayu Chatterjee, Daniel Parker, Junkai Dong, Tomohiro Soejima, and Mit Naik for helpful discussions. 
This work was primarily funded by the U.S. Department of Energy, Office of Science, Office of Basic Energy Sciences, Materials Sciences and Engineering Division under Contract No. DE-AC02-05-CH11231 (Theory of Materials program KC2301) (TW, WK, SL, MZ). 
LF was supported by the Air Force Office of Scientific Research (AFOSR) under award FA9550-22-1-0432.
This research used computational resources from the Lawrencium cluster provided by the IT Division at the Lawrence Berkeley National Laboratory, the National Energy Research Scientific Computing Center (NERSC) (both supported by the Office of Science, the U.S. Department of Energy under Contract No. DE-AC02-05CH11231), and Frontera at the Texas Advanced Computing Center (TACC) (supported by the National Science Foundation under grant No. OAC-1818253).

\end{acknowledgments}

\bibliography{main, CFL_bib, bibliography_TBG, bibliography}




 


\end{document}